\documentclass[12pt,epsf] {article}
\usepackage{amsfonts}
\begin{document}

\newcommand{\BOX}{\hfill $\Box$}
\newcommand{\NAB}{\hfill $\nabla \nabla \nabla$}
\newcommand{\BYDEF}{\stackrel{\Delta}{=}}

\newcommand {\beq} {\begin{equation}}
\newcommand {\eeq} {\end{equation}}
\newcommand {\bear} {\begin{eqnarray}}
\newcommand {\eear} {\end{eqnarray}}
\newcommand {\bears} {\begin{eqnarray*}}
\newcommand {\eears} {\end{eqnarray*}}
\newcommand {\done} {\quad\vrule height4pt WIDTH4PT}
\newcommand {\phantomI} {\vrule height20pt WIDTH0PT}
\newcommand {\barr} {\begin{array}}
\newcommand {\earr} {\end{array}}
\newcommand {\N} {\rm I\!N}

\def\Ooo{{ \Omega}_0^0}
\def\Oooc{{ \Omega}_0^{0,c}}

\def\bmr{\, \in \mbox{\boldmath $R$}}
                \def\dvl{\|_{2}^{2}}
                \def\hi{H_{\infty}}
                \def\eq{equation}
                \def\eqq{eqnarray}
                \def\eqn{eqnarray*}
                \def\lab{\label}
                \def\rn{{\bf R}^{n}}
                 \def\rmm{{\bf R}^m}
                 \def\rk{{\bf R}^k}
                   \def\rnm{{\bf R}^{n\times m}}
                    \def\rnk{{\bf R}^{n\times k}}
                    \def\ep{\epsilon}

\def\limi{\mathop{\underline{\rm lim}}}
\def\star{*}
\def\cL{{\cal L}}
\def\cP{{\cal P}}
\def\tL{\tilde L}
\def\ztla{z_{\tau_l}^1}
\def\ztlb{z_{\tau_l}^2}
\def\intt{\int_{\tau_l}^{\tau_{l+1}}}
\def\summ{\sum_{n=\lf{\tau_l /\epsilon} + 1 }^{
\lf{\tau_l /\epsilon} + K(\epsilon ) } }
\def\cZ{{\cal Z}}
\def\ue{u_\epsilon}
\def\uy{{\ue(y)}}
\def\Qz{{\bf Q_0}}
\def\Qe{{\bf Q_\epsilon }}
\def\lf#1{{\lfloor #1 \rfloor}}
\def\lfe{{\lfloor \epsilon^{-1} \rfloor}}
\def\lims{\mathop{\overline{\rm lim}}}
\def\dfn{ \stackrel{\rm def}{=} }
\def\sp{\newline\noindent}
\def\sqr#1#2{{\vcenter{\hrule height.#2pt
      \hbox{\vrule width.#2pt height#1pt \kern#1pt
        \vrule width.#2pt}
    \hrule height.2pt}}}
\def\square{\vrule height6pt width7pt depth1pt}
\def\endpf{\hfill\square\bigskip}
\def\R{\mathop{\rm I\kern -0.20em R}\nolimits}
\def\Prf{\noindent{\bf Proof: }}
\def\fA{{\bf A}}
\def\fH{{\bf H}}
\def\ty{\tilde y^\epsilon}
\def\tz{\tilde z^\epsilon}
\def\fX{{\bf X}}
\def\cF{{\cal F}}
\def\cS{{\cal S}}
\def\tu{\tilde u}
\def\ts{\tilde s}
\def\cM{{\cal M}}
\def\fY{{\bf Y}}
\def\bV{{\bf Y}}
\def\R{\mathop{\rm I\kern -0.20em R}\nolimits}
\def\G{{\bf G}(g)}       \def\F{{\cal F}}
\def\C{{\cal C}}         \def\D{{\cal D}}
\def\nn#1{\mathop{\left |\kern -0.10em \left | #1
        \right |\kern -0.10em \right |} \nolimits}
\def\pd#1#2{\frac{\partial #1}{\partial #2}}

\def\uez{u_{\ep}\left(\bar{z}(t)\right)}
\def\exu{E_x^{u_{\ep}(\bar{z})}}
\def\dee{\Delta(\ep)}
\def\idee{\frac{1}{\dee}}
\def\tle{{\lfloor{\tau_l}/{\epsilon}\rfloor}}
\def\tl1e{{\lfloor{\tau_{l+1}}/{\epsilon}\rfloor}}

\newtheorem{definition}{Definition}[section]
\newtheorem{algorithm}{Algorithm}[section]
\newtheorem{lemma}{Lemma}[section]
\newtheorem{corollary}{Corollary}[section]
\newtheorem{proposition}{Proposition}[section]
\newtheorem{theorem}{Theorem}[section]
\newtheorem{remark}{Remark}[section]
\newtheorem{example}{Example}[section]
\newtheorem{assumption}{Assumption}[section]
\newtheorem{conjecture}{Conjecture}[section]
\newtheorem{postulate}{Postulate}[section]
\newtheorem{hypothesis}{Hypothesis}[section]

\makeatletter
\def\EquationsBySection{\def\theequation{\thesection.\arabic{equation}}%
\@addtoreset{equation}{section}}
\makeatother
\EquationsBySection

\title{{\bf Coupling Control and Human-Centered Automation in Mathematical Models
of Complex Systems}}

\author{
{\large \hspace{-0cm} Roderick V. N. Melnik\thanks{Tel.: +1-519-884-1970 (3662);
E-mail: rmelnik$@$wlu.ca }
}\\
{\large \hspace{-0cm} Mathematical Modelling \& Computational Sciences,}\\
{\large \hspace{-0cm} Wilfrid Laurier University, Waterloo Campus,}\\
{\large \hspace{0cm}  75 University Avenue West, Waterloo, ON, Canada N2L 3C5 } \\
}

\date{}
\maketitle
\thispagestyle{empty}

{
\abstract{It is known that with the increasing complexity of technological systems that
operate in dynamically changing environments
 and require human supervision or a human operator,  the relative share of human errors
is increasing across all modern applications. This indicates that in the analysis
 and control of such systems,
human factors should not be eliminated by conventional formal mathematical
methodologies. Instead, they must be incorporated into the modelling framework giving rise
to an innovative concept of human-centered automation.

In this paper we analyze mathematically how such factors can be effectively incorporated
 into the analysis and control of complex systems. As an example,
we focus our discussion around one of
the key problems in the Intelligent Transportation Systems (ITS)
 theory and practice, the problem of speed control, considered
here as a decision making process with limited information available. The problem is cast
mathematically in the general framework of control problems and is treated in the context of
dynamically changing environments where control is coupled to human-centered automation.
Since in this case   control might not be limited to a small number of control settings, as
it is often assumed in the control literature, serious
 difficulties arise in the solution of this problem. We demonstrate that
the problem can be reduced to a set of Hamilton-Jacobi-Bellman
equations where human factors are incorporated via estimations of the system Hamiltonian.
 In the ITS context,  these estimations can be obtained with the use
of on-board equipment like sensors/receivers/actuators, in-vehicle
communication devices, etc. The proposed methodology provides a way to integrate human factor
 into the solving process of the models for other complex dynamic systems. }

}

\bigskip
\noindent
{\bf Key words:} complex dynamic systems,
 artificial intelligence, speed control, intelligent
transportation systems, human factors, Hamiltonian estimations.

\bigskip

\section{Introduction}
\vspace*{0.3cm}

It is generally accepted that much of human intelligence can be characterized as the ability
to recognize complex patterns, to analyze them and, if possible, to control.  In this process
the visual system, among others, together with cognition play a central role
 (\cite{Carroll2003}, p.11). In creating advanced technological
 systems human factors modelling must be incorporated as the processes of complex
pattern recognition,  their analysis, and ultimately control are intrinsically hierarchical.
In this contribution we demonstrate how this can be achieved on an important example from the
ITS theory - the problem of speed control. The main reason for this choice lies with the fact
that while being strongly dependent on human factors, efficient speed control is known to be
one of the key problems in the ITS technology \cite{Endo1999A,Endo1999B,Endo2000,Seto1999}.
For the purpose of this paper we limit ourselves to three main technological analogies of
human intelligence mentioned already, pattern recognition, analysis, and control. In the
context of ITS technology such analogies
 are pertinent
to (a) the application of information-driven functions (software for both control and computation)
and (b) communications
systems to controlling traffic (i.e. operating transport effectively, handling
emergencies and incidents if they arise, automating driving and safety, etc).
These aspects are in the heart of the development of intelligent vehicles and
highway systems (e.g., \cite{Gollu1998} and references therein).

Having specified our focus area from where all our examples will be drawn,
we note further that our discussion will be
pertinent to mathematical models for the development of automated driving
strategies based on a regulated speed control. Such strategies are important
in many areas including  collision avoidance (e.g.,
 control the vehicle with respect to a vehicle running ahead, control the merging
process into a main traffic steam where the ``target'' vehicle is
running), the minimisation of the fuel
 consumption, etc.
Under the requirements of increased safety, minimisation of the fuel consumption,
and strict environmental constraints imposed by the government,
 many automotive and transport engineering companies have increased their
attention to this problem
\cite{Endo1999A,Endo1999B,Stotsky1995,Endo2000,Seto1999,Butts1999,Fontaras2007, Manzie2007}.
The increased complexity of intelligent transportation systems in this area and the
successful development of automated driving strategies require accounting for human-related
design factors and the integration of these factors into mathematical models used. These
factors remain an important link in a chain of automated driving strategies developed from
the application of mathematical models. Although there is no general model describing the
dynamics of human interaction with complex systems in dynamically changing environment
\cite{Rouse1993,Goodrich2000}, by analyzing existing approaches applied previously to some
model transport problems, in this paper we suggest  a simple and efficient way to account for
human factors in the solution of the speed control problem by considering a sequential
HJB-equation-based approximation of the system Hamiltonian. Human-centered technologies are
used frequently in many
 applications, including artificial intelligence \cite{Shahar2006,Wren2006}.
As pointed out in \cite{Kesseler2006}, although much system development is still currently
done by using a technology-centered approach (that is automating the functions the technology
is able to perform), we witness an increasing-in-importance
 potential of human-centered design where we combine skilled human and automated support.
 This relatively new paradigm, has already demonstrated its importance in complex
system development where intervention of humans is still necessary on supervisory basis
and/or at certain stages of system evolution (e.g., \cite{Mayer2006} and references therein).
Nevertheless, the body of literature in this area is minimal
\cite{Barthelemy2002,Kraiss2001,Mayer2006}, let alone mathematically rigorous developments.

From a methodological point of view the approach we develop in this contribution
can be viewed
as a blend of control and human-centered automation aspects in the design/control of
intelligent transportation  systems where we have to satisfy
 often competing requirements of human and technological objectives accounting
for their capability limitations and constraints \cite{Goodrich2000}.
 The proposed approach is generic enough to be applicable to system developments
in application areas outside of the ITS domain. Finally, we note that our approach has much
in common with the paradigm of supervisory control \cite{Kirlik1993,Melnik1997} where, in the
context of our problems, the control algorithm should respond in real time to changing
conditions where the underlying process can be represented  in a space of discrete events
\cite{Melnik1998}. Taking this point of view into account, we structure the rest of this
paper as follows.

\noindent $\bullet$ In Section II we provide a general mathematical framework for controlling
complex systems by using continuous and discrete control settings. The discussion is given in
the context of the ITS speed control problem.

\noindent $\bullet$ In Section III we consider a specific example of the speed control
problem subjected to the minimization of the fuel consumption.

\noindent $\bullet$ In Section IV  two important approaches to the development of automated
driving strategies are discussed and difficulties in their computational implementation are
analyzed in detail with
 exemplification given for the problem considered in Section III. In this section it is also
 shown that the general speed control problem can be reduced to a model based on the solution
 of HJB-type equations
where  human factors are incorporated naturally via estimations of the system Hamiltonian.

\noindent $\bullet$ Concluding remarks  are given in Section V.

\section{Mathematical formulation of the problem,
 exemplified for advanced vehicle system control}

While the formulation given in this section can be easily adapted to control of other complex
systems, we exemplify our discussion here with an example concerning control of advanced
vehicle systems. The model-based computer-aided control has become an intrinsic component of
 ITS technology. Due to the increased
complexity and tight coupling of many different constraints imposed on the automotive systems
development process, this control becomes increasingly important \cite{Sivashankar1999}. Such
constraints come from the growing environmental and economic concerns leading to the rising
customer expectations for fuel economy, performance, tightening emission, etc. There is a
growing expectation that these constraints could be resolved by developing advanced
transportation technologies \cite{Kolmanovsky2000}. Since many of these constraints are
dependent strongly on the choice of driving strategies, this leads to the necessity of
coupling control with human-centered automation at the level of modelling rather than at the
stage of system utilization. Human-centered automation is a relatively new concept that plays
an increasingly important role in new technologies, both military and civil
\cite{Goodrich2000,Kraiss2001}. While computers are in the heart of control of most complex
man-designed systems, full automation is often either not feasible or not reasonable, in
particular if system and/or environment conditions are rapidly changing. In situations like
this, function allocation and coupling between human factors and automation become critical
\cite{Kraiss2001}. In this contribution, we demonstrate that this coupling can be treated
formally via a  sequential estimation of the system Hamiltonian, providing an important tool
in theory and applications of ITS and other complex systems.

First, we formulate the problem of interest in the general framework of control problems
providing all the explanations on an example of the analysis of the situation on the road
during the interval time $[0,T]$, followed by the subsequent development of automated control
strategies for road participants (driver-vehicle subsystems). This problem can be formulated
as minimization of the following functional
\begin{eqnarray}
J({\bf u}, {\bf v}) = \int_0^T  f_0({\bf x}(t), t, {\bf v}(t), {\bf u}(t)) d t \rightarrow \min,
\label{eq1}
\end{eqnarray}
where ${\bf v}, {\bf u}$ are ${\mathbb{R}}^m \rightarrow {\mathbb{R}}^m$
functions that represent the velocity and control of the entire dynamic system consisting of $m$ subsystems  ($m \in \mathbb{N}$, e.g. the number of vehicles), and the function $f_0$ is the objective (problem-specific) function that could characterise fuel consumption, emissions, etc (or a combinations of those quantities
 incorporated via corresponding weights). In (\ref{eq1}) $T$ is
the prescribed (or estimated maximum) time, e.g. the time of
reaching the final destination,  $\displaystyle {\bf v} = \frac{ d\; {\bf x} }{ d \; t}$, ${\bf x}(t) =(x_1, x_2,...,x_m) \in {\mathbb{R}}^m$ is the position of the road participants at time $t$ with applied (speed) control  ${\bf u}$.
The dynamics of coupling between the velocity and control is governed not only by (\ref{eq1}) but also by the state constraints that are assumed to have the form of the equation of motion
\begin{eqnarray}
\frac{ d {\bf x}}{d t} = {\bf f} ({\bf x}, t, {\bf u}(t))
\label{eq2}
\end{eqnarray}
and/or second Newton's law
\begin{eqnarray}
\frac{ d {\bf v}}{d t} = {\bf F} ({\bf x}, t, {\bf v(t)}, {\bf u}(t)).
\label{eq3}
\end{eqnarray}
 The vector functions ${\bf f}$ and ${\bf F}$ can be viewed as problem-specific
 approximations to the velocity function and acceleration (see e.g. \cite{Howlett1990,Stotsky1995}
  for some specific examples).

The qualitative (and quantitative) behaviour  of the solution of this problem will be
determined at a large extent by control constraints
 \cite{Melnik1997}, defined here as
\begin{eqnarray}
{\bf u}(\cdot) \in {\cal U}(t, {\bf x}),
\label{eq4}
\end{eqnarray}
where  ${\cal U}(t, {\bf x})$ is a given space-time set. It should be noted that for the
solution of the speed control problem for transport
 engineering systems both groups of models,
 with continuous control and with discrete control, have been used in the literature (see, e.g., \cite{Howlett1995} and references therein). Although models with continuous control have limited applicability in this context (moreover,
their analysis typically requires the assumption of a finite number of control settings
\cite{Howlett1990}), they provide an important insight into more realistic models with
discrete control. A major difficulty with the existing approaches based on continuous control
models becomes transparent at the computational level where the quality of results depends
heavily on the number and the form of {\em a priori} chosen control (traction) phases (e.g.
power, coast, brake).  A similar difficulty exists for models with discrete control settings,
where the total number and locations of "switching control points"  largely determines the
quality of computational results. More precisely, the ``switching control point'' problem can
be reduced to the determination of ``optimal'' times
\begin{eqnarray}
0=t_0 < t_1 <...<t_{n}
\label{eq5}
\end{eqnarray}
which correspond to such a partition of the vehicle trajectories  ${\bf x}^k={\bf x}(t_k),$
$k=0,1,...,n$ that control at those points makes the entire trip optimal in some specified
sense. In conventional approaches, the responsibility on choosing the precise sequence of
control settings and the determination of the ``optimal'' positions of these switching points
can implicitly be shifted to the driver \cite{Howlett1996}.  However, this becomes
undesirable in the context of ITS where the driving strategy should be automated effectively
to minimise the probability of accidents and to satisfy other goals of traffic control. Since
the performance of the entire dynamic system can be improved greatly by increasing the number
of discrete control settings, the ITS technology can provide an effective way to achieve
these goals by implementing highly efficient driving control strategies
 on automated highway systems (AHS), a next generation of road systems that are intended
 to resolve various traffic issues \cite{Seto1999}.
Such driving strategies can be developed from the solution of problem (\ref{eq1})--(\ref{eq4})
for a sufficiently high number $n$ in (\ref{eq5}).
The practical implementation of such strategies for large $n$ will require the installation of
on-board equipment such as actuators for controlling the breaks and throttle,
 LCX (leakage coaxial cable) receivers, as well as
a laser radar and inter-vehicle communication devices. Then, in principle, the vehicles can be
operated according to  a vehicle velocity command
(the indicated vehicle speed, road grade, road curvature, and accident information) received
from the LCX cable installed alongside the road, that allows for automatically maintaining a
 safe vehicle speed and headway distance.
We emphasize that in this case, the definition of switching times (\ref{eq5}) will be made
sequentially on the basis of information accumulated by the given moment of time, as opposed
to the conventional techniques based on one of the {\em a priori} choices of switching times.

An example involving one vehicle only is discussed in the next section in order to clarify
the meaning of functions in general control problem (\ref{eq1})--(\ref{eq5}) and to lay the
foundation for further discussion of the key issues related to the solution of this problem
in the context of ITS.

\section{Conventional approaches on the example of
vehicle speed control subjected to minimization of fuel consumption}

The literature on different aspects of control of transportation systems is vast (e.g.,
\cite{Howlett1995,Kolmanovsky2000,Qi2005,Zhuan2006} to name just a few). A number of authors
have attempted to apply different variants of continuity principle to determine switching
control times (where, e.g., intervention of the driver is required). A continuity hypothesis
found also its application in continuum (fluid-dynamics-like) approaches that have been
developed for traffic flow models. In the latter case,  such models rely frequently on
unrealistic sets of assumptions and an {\em a priori} optimal velocity is one of them. More
recently, several interesting contributions have been made to this area where authors
realized that the underlying problem can be modelled with a hyperbolic system (with no
conservation of momentum, e.g. \cite{Herty2006} and references therein). However, the authors
of these recent papers do not discuss control issues and that is where major challenges lie.

Let us explain the situation on an example of vehicle speed control subjected to the
minimization of the fuel consumption. First note that in a number of practical situations the
general formulation of problem (\ref{eq1})--(\ref{eq4}) can be simplified considerably by
assuming that control and state aspects of the dynamics of the moving vehicle could be
decomposed (or factorized) in the objective function, i.e. if we assume  that
\begin{eqnarray}
f_0({\bf x},t, {\bf u}) = p[{\bf u}(t)] \cdot q[{\bf v}(t)],
\label{eq6}
\end{eqnarray}
where $p$ and $q$ are given functions. For example, according to \cite{Howlett1990},  for a
problem where the total mechanical energy consumed by the vehicle is given by  (\ref{eq1})
and (\ref{eq6}) and  control is subject to the minimization of the fuel consumption, the
above functions can be defined in the following forms (note that $m$ is set to 1 in this
case)
\begin{eqnarray}
p \equiv u_+(t) = \frac{1}{2} [ u(t) + |u(t)| ], \quad q \equiv v(t).
\label{eq7}
\end{eqnarray}
As pointed out in \cite{Howlett1990}, Eq. (\ref{eq7}) makes sense when a maximum applied
acceleration is specified and that only positive acceleration consumes energy. We note
further that specific forms of constraints (\ref{eq2}) and/or (\ref{eq3}) depend on the
nature of the problem at hand, and since the vehicle dynamics can be influenced by the
engine, automatic transmission, breaks and by many other factors, the constraints can appear
to be fairly complex in the general case. Nevertheless, for a number of important situations
state constraints can be reasonably simplified. For example, it is often assumed that the
control variable is the applied acceleration and that this variable can be determined as the
difference between  the ``controlled'' acceleration function  $s(u,v)$ (from a physical point
of view this function can be interpreted as the driving controlled force per unit mass of the
vehicle) and the ``uncontrolled'' deceleration function $r(v)$ of the vehicle. In this case
equation (\ref{eq3}) takes the form
\begin{eqnarray}
\frac{d v}{d t} = s[x, u(t),v(t)] -r [x,v(t)], \label{eq8}
\end{eqnarray}
where possible dependency of functions $s$ and $r$ on the position $x$ has been also
included.  This can be simplified further. Note that the form of the deceleration is again
problem-specific depending on the need to account for a number of factors such as
contributions of gravitational, aerodynamic, frictional and other forces. In the simplest
case it can be approximated by the difference between the frictional resistance, $r_0$, and
the gravitational component $g$ in the direction of the vehicle motion \cite{Howlett1996}:
\begin{eqnarray}
r[v(t)]= r_0[v(t)]- g(x). \label{eq9}
\end{eqnarray}
Note also that (\ref{eq9}) often takes the form of a quadratic law (so-called Davis' formula)
\begin{eqnarray}
r[v(t)] = a + b v + c v^2 \quad a,b,c \in \mathbb{R}.
\label{eq10}
\end{eqnarray}
It is often the case that additional inequality constraints  come naturally into the formulation of the problem. For example,  the definition of the control variable might require further constraints such as positivity of the velocity and  some control admissibility conditions (e.g. \cite{Howlett1990})
\begin{eqnarray}
v(t) \geq 0, \quad
|u(t)| \leq 1.
\label{eq11}
\end{eqnarray}
These constraints can be supplemented by additional constraints such as an upper limit on
velocity. Inequality constraints (\ref{eq11}) can be cast in the general vector form as
\begin{eqnarray}
{\bf G} \leq {\bf 0}, \quad \mbox{with} \quad G_1=-v, \quad G_2 = u^2-1.
\label{eq12}
\end{eqnarray}
Furthermore, some equality constraints might be also required. For example,
if we assume that the trip has  length $L$, this leads to the equality constraint expressed by the end-point reachability condition
\begin{eqnarray}
\int_0^T v(t) d t =L,
\label{eq13}
\end{eqnarray}
supplemented by the boundary conditions
\begin{eqnarray}
v(0)= v_1, \quad v(T)=v_2
\label{eq14}
\end{eqnarray}
taken typically with $v_1= v_2=0$.

Next, we note that the above example can be reformulated in the general framework
(\ref{eq1})--(\ref{eq4}) by introducing vector ${\bf x}=(x_1, x_2)^t$ with $x_1$ being the
state variable, and $x_2$ being the velocity of the vehicle. Indeed, since $\displaystyle
\frac{d x_1}{d t} =v$, we can use (\ref{eq14}) to derive that
\begin{eqnarray}
x_1(T) - x_1(0) =L.
\label{eq15}
\end{eqnarray}
Then, taking into account (\ref{eq14}) we have the initial and terminating conditions in the form
\begin{eqnarray}
{\bf x}(0) = {\bf 0}, \quad {\bf x} (T) = {\bf x}_T,
\label{eq16}
\end{eqnarray}
where ${\bf x}_T =(L,0)^t$.  We denote
\begin{eqnarray}
{\bf f} ( {\bf x}, t, {\bf u} (t)) = (x_2, s[x_1, x_2, u_2] - r[x_1, x_2])^t,
\label{eq17}
\end{eqnarray}
where $u_2$ plays the role of $u$ in the above example,  ${\bf u}=(0, u_2(t))^t$, and take
into account (\ref{eq12}) and (\ref{eq16}), i.e. only admit controls
\begin{eqnarray}
{\bf u}(\cdot) \in {\cal U}(t, {\bf x}),
\label{eq18}
\end{eqnarray}
where
\begin{eqnarray}
{\cal U} (t, {\bf x}) = \left \{ {\bf u} (\cdot) \in {\cal U}^0(t):  x_2(t) \geq 0, \;
\right.
\nonumber
\\[10pt]
\left. {\bf x}(T) = x_T, \; |u_2(t)| \leq 1 \right \}. \label{eq19}
\end{eqnarray}
Then, the definition of the objective function in the form (see (\ref{eq6}))
\begin{eqnarray}
f_0({\bf x}(t), t, {\bf u}(t)) = [u_2(t)]_+ x_2(t)
\label{eq20}
\end{eqnarray}
completes the formulation of the vehicle speed control problem subjected to the minimization
of the fuel consumption in the general framework (\ref{eq1})--(\ref{eq4}).

Now, we are in a position to highlight major difficulties in applications of conventional
methodologies to the above problem. First, we note that in reality the control variable of
this problem cannot vary continuously due to the discrete character of the information
\cite{Melnik1998} obtained in this specific case by the moving vehicle in a dynamically
changing environment. Therefore, if we consider a finite (possibly very large) set of control
settings, for example, throttle settings as it was originally proposed in \cite{Howlett1990}
\begin{eqnarray}
-1=u^1 < u^2 < ... < u^n =1, \label{eq21}
\end{eqnarray}
then the analysis of the problem can be reduced (under some severe assumptions such as "no
speed limits") to the consideration of four basic situations, as shown in \cite{Howlett1990}
(the acceleration, speedholding, coasting, and breaking phases), making use of quite specific
forms of functions (\ref{eq7}). In such cases it might be easier to define the objective
function of the total fuel consumption accounting for these settings by splitting the total
distance on an appropriate number of sub-intervals according to the discrete dynamic
 equation $x^k=x(t_k)$ with
\begin{eqnarray}
0=x^0 <x^1 < ... < x^n =X
\label{eq22}
\end{eqnarray}
and by assuming that the time ${\Delta t}_{i+1} =t_{i+1} -t_i$ required to complete the
segment trip $[x_{i}, x_{i+1}]$ is known (or can be well approximated) for all
$i=0,1,...,n-1$ (see (\ref{eq5})). In most conventional approaches referenced here it is
assumed that each control setting determines a constant rate supply. If we denote the the
fuel consumption and the control setting in the interval  $[x_{i}, x_{i+1}]$ by $c_{i+1}
\equiv c[u^{i+1}]$ and $u^{i+1}$ ($c=0$ if $u \leq 0$), respectively, the cost (fuel
consumption) functional of the entire trip can be defined as \cite{Pudney1994,Howlett1996}
\begin{eqnarray}
J= \sum_{i=0}^{n-1} c_{i+1} {\Delta t}_{i+1}. \label{eq23}
\end{eqnarray}
  First observe that since in the general case all control settings $c_{i+1},$
$i=0,1,...,n-1$ are functions of time, a more rigorous approach should be based on the
consideration of functional (\ref{eq1}) rather than function (\ref{eq23}). We observe also
that in some cases (including more realistic situations with speed limits), the development
of automated (optimal or sub-optimal) driving strategies can be reduced to the analysis of
simple combinations between a small number of control settings (e.g., power when $u=1$, coast
when $u=0$, and break when $u=-1$). However, due to the very nature of the control problem
where we have to consider the ITS in a dynamically changing environment,  a more detailed
{\em sequential} analysis of the whole information sequence (\ref{eq21}) is required. Such an
analysis is intrinsic to other control problems where complex systems require human
supervision or a human operator. The proposed methodology for this analysis is discussed in
the next section.


\section{Sequential analysis of the global Hamiltonian keeps the key to efficient driving strategies}

In the context of our example, the information sequence for the decision making process
obtained by on-board LCX receivers and by inter-vehicle communication devices always contains
a certain degree of uncertainty. Indeed, some of the vehicle parameters, as well as the
information on road conditions, can be known only partially \cite{Stotsky1995}. A complex
dynamics of human performance in traffic systems \cite{Rouse1993} brings along
 another factor that complicates the analysis of the entire dynamic system consisting of many driver-vehicle subsystems. This leads to a situation where control cannot be limited by a simple combinations of basic settings, as we have discussed in the previous section, and a general approach should be developed to address the speed control problem in the ITS context.

The development of speed control strategies for intelligent transportation systems has become
an important and topic of research \cite{Stotsky1995,Kiencke2006}. In this section, our
discussion focuses on a subset of the systems that consist of
  vehicles  capable of measuring/estimating dynamic information from the target (typically, the immediate front) vehicle by its on-board sensors.
The computers in the vehicles can process the measured data and generate proper throttling
and breaking actions for controlling vehicles' movements under the constraints of safety,
ride comfort, fuel minimization, etc. Recall that algorithms for speed control with constant
acceleration/deceleration were developed and tested together with some simple algorithms for
``approach'' and ``merging'' control  \cite{Endo1999A}. The authors of \cite{Endo1999A}
developed linear models for passenger cars (in which any acceleration/deceleration can be
generated according to the driver's operations \cite{Endo1999B}) and generalize their results
to a nonlinear model for heavy-duty vehicles where they account for transient responses (it
is rather difficult to control the speed in this case, because of poor acceleration
performance of such vehicles). As it was shown, it is necessary to account for
saturation/delay in acceleration which could be an important characteristic of some vehicles.
However, the results of simulations conducted in \cite{Endo1999A} showed that for long
control periods, the model leads to unrealistic speeds, exceeding  the target speed, and the
maximum vehicle distance could become excessively long. In principle, such overshootings can
be avoided by setting a short control period. However, since the dynamic behaviour of the
vehicle model is intrinsically nonlinear and considerably complicated  \cite{Stotsky1995}, in
the general case it is necessary to take into account complex dependency between acceleration
and speed of the vehicle using the general framework of (\ref{eq1})--(\ref{eq4}).

In the reminder of this section we analyse three main approaches to the solution of the general speed control problem with exemplification given for
  the vehicle speed control subjected to the minimization of the fuel consumption, as considered in Section III.

\subsection{The definition of the Hamiltonian via the solution of the adjoint problem}

Some of the most powerful methodologies to the analysis of speed control problem are based on a heuristic application of the Pontryagin maximum principle.  However, the application  of these methodologies
to solving practical problems in the context of intelligent transportation systems requires overcoming a number of serious difficulties
which will be considered below in the case where ${\bf x} \in {\mathbb{R}}^2$ (see details after (\ref{eq14}) in Section III).

Applying formally the Pontryagin maximum principle \cite{Pontryagin1986} to the problem considered in Section III (problem (\ref{eq1})--(\ref{eq4}) allows an analogous treatment), we can introduce a local Hamiltonian of the entire dynamic system in the following form
\begin{eqnarray}
H({\bf x}(t), {\bf u}(t),  \vec\psi(t), t) = - a_0 f_0({\bf x}(t), {\bf u}(t), t) + \sum_{i=1}^2 \psi_i f_i,
\label{eq24}
\end{eqnarray}
where all notations come from the consideration of (\ref{eq1})--(\ref{eq4}) in this special
case,  ${\bf f}=(f_1, f_2)^T$, $a_0$ is the normalisation factor \cite{Melnik1997},  while
the adjoint vector-function $ \vec\psi=(\psi_1, \psi_2)^T$ is defined from the following
adjoint system
\begin{eqnarray}
\frac{\partial \psi_i}{\partial t} = \frac{\partial f_0}{\partial x_i} - \sum_{k=1}^2 \psi_k \frac{\partial f_k}{\partial x_i}, \quad \mbox{where} \; \; \psi_i(T)=0, \; \; i=1,2.
\label{eq25}
\end{eqnarray}
Then the result of the application of the Pontryagin maximum principle to the speed control problem
can be formulated as follows \cite{Pontryagin1986}.
\begin{theorem}
For a driving strategy determined by the pair $({\bf u}(t), {\bf x}(t))$ to be optimal it is
necessary the existence of an adjoint vector-function $\vec\psi(t)$ (components of which are
not identical zero), defined by (\ref{eq25}) such that
\begin{eqnarray}
\max_{u \in {\cal U}} H(\vec\psi(t), {\bf x}(t), {\bf u}(t), t) =
H({\vec\psi}^*(t), {\bf x}^*(t), {\bf u}^*(t), t)
\label{eq26}
\end{eqnarray}
for almost all $t \in [0, T]$.
\end{theorem}
Practical difficulties with the application of this approach to the solution of the speed
control  problem for the intelligent transportation systems lie with  the fact that state
variables in this  problem are not independent. This fact has led many authors to substantial
simplifications of the problem (in particular, in the analysis of the system Hamiltonian) by
considering a small subset of possible control settings \cite{Howlett1995}. Unfortunately,
this idea cannot be applied in the context of intelligent transportation systems, because
both state variables are closely linked with the control function ${\bf u}$, and they might
be decoupled in some special situations only.

Consider, as an example, the problem  discussed in Section III. In this case, functions
participating in the definition of Hamiltonian (\ref{eq24}) can be specified more precisely.
Indeed, recall that in this case function $f_0$ takes the form (\ref{eq20}), while the vector
function ${\bf f}$ can be specified by its components as in (\ref{eq17}). Clearly that even
in this relatively simple case the state variables are coupled with the control by the
following systems of equations
\begin{eqnarray}
\frac{\partial x_1}{\partial t} = x_2, \quad \frac{\partial x_2}{\partial t} = s[x_1(t), x_2(t), u_2(t)] - r[x_1(t), x_2(t)]
\label{eq27}
\end{eqnarray}
supplemented by  the corresponding boundary conditions and other constraints previously
discussed. In this case, according to (\ref{eq24})  the Hamiltonian of the system  can
formally be written in the form
\begin{eqnarray}
H= - a_0 {[u_2]}_+ x_2 + \psi_1 x_2 + \psi_2 \left \{ s[x_1, x_2, t, u_2]-
r[x_1, x_2] \right \}.
\label{eq28}
\end{eqnarray}
For example, in a special case   where $s[x_1, x_2, t, u_2]=u_2$ and $a_0=1$
\cite{Howlett1995},   control can be confined to the following values $u_2=1$, $u_2 \in
(0,1)$, $u_2=0$, $u_2 \in (-1,0)$ and $u_2=-1$ subject to the fulfilment of one of the
following five relations (a) $\psi_2 > x_2$, (b) $\psi_2= x_2$, (c) $0 < \psi_2 < x_2$, (d)
$\psi_2=0$, (e) $\psi_2 <0$, respectively. Such a consideration takes the advantage of an
implicit assumption on the possibility of decoupling  state and control aspects of the
problem. This leads to a substantial simplification of the analysis where we have to account
for control constraints (\ref{eq4}). In the general case, the analysis cannot be reduced to
five basic situations described above. Since the intersection between the set defined by
control constraints and the set defined by state constraints is not empty \cite{Melnik1997},
we note that even under these special assumptions, the key to the analysis of the Hamiltonian
 is kept by the coupled system of equations (\ref{eq25}), (\ref{eq27}). Indeed, the adjoint
function of the speed control problem considered in Section III is the solution of
(\ref{eq25}) which in the case where $s$ equals $u_2$ can be written in the following form
\begin{eqnarray}
\frac{d \psi_2(t)}{d t} - r^{\prime} [v_0(t)] \psi_2 = \tilde{F},
\label{eq29}
\end{eqnarray}
where
\begin{eqnarray}
\tilde{F} \equiv \tilde{F}(v_0, u_0, f_0, \tilde{f}_2, \tilde{c}),
\label{eq30}
\end{eqnarray}
 $\tilde{f}_2: \; \mathbb{R} \rightarrow \mathbb{R}$ is a real function associated with
dynamically perturbed function $F_2=\displaystyle \frac{\partial f_2} {\partial x_2}$, and
$\tilde{a}_0 \in \mathbb{R}$ is a real constant that can be viewed as a dynamically perturbed
parameter of normalization, subject to the dynamics of $\psi_1$. Getting a specific form of
$\tilde{F}$ requires a quite delicate analysis which was performed so far only for fairly
simple cases (e.g. \cite{Howlett1990,Howlett1995} and references therein). The determination
of function $\tilde{f}_2$ and constant
 $\tilde{a}_0$ is also far from trivial
 and in the general case  such a determination
should be {\em adaptive}.  Note also that  $(v_0, u_0)$ in (\ref{eq30}) is   assumed to be a
fixed point associated with the optimal velocity of the vehicle and its optimal applied
control which are not known {\em a priori}. However, if an approximate  solution of the
problem (\ref{eq29}) is found,  then by using  (\ref{eq24}) a local (or pointwise)
 Hamiltonian function of the system can be defined.
In this case, a major source of difficulties in constructive approximations of  optimal
driving strategies (that can be derived formally from minimising the local Hamiltonian) lies
with the intrinsically complex dynamics of the adjoint function and the adequate
determination of the normalisation factor. To proceed with such a construction the local
Hamiltonian function should be integrated in time over the whole interval  $[0, T]$ which
leads to the global Hamiltonian in the form
\begin{eqnarray}
{\cal H}(u) = \tilde{H}_0 + \int_0^T H[x_1, x_2, t, u_2, \psi_1, \psi_2] d t,
\label{eq32}
\end{eqnarray}
where the actual value of $\tilde{H}_0 \in \mathbb{R}$ depends not only on $v_0$, $L$, and
$T$, but also on the  weight coefficients for implementing all remaining constraints of the
problem. The optimal control strategies can now be determined by finding local minima of the
Hamiltonian (\ref{eq32}), but in practice this
 approach leads to serious computational difficulties due to too many degrees of freedom in
(\ref{eq32}). On the other hand, the problem can be reduced to the analysis of (\ref{eq24}),
e.g. by considering a small subset of basic control settings, only in quite simple cases such
as those discussed in \cite{Howlett1990}.

\subsection{Using the embedding principle and the Lagrangian multipliers}

In the context of intelligent transportation systems, more  feasible computationally are
approaches that are based on the
 embedding principle.
First, we introduce the minimum cost function as follows
\begin{eqnarray}
J^*({\bf x}(t), t) = \min_{{\bf u}(\tau) \in {\cal U}, \; t \leq \tau \leq T}
\left \{ \int_t^T f_0 ({\bf x}(\tau), {\bf u}(\tau), \tau) d \tau \right \},
\label{eq33}
\end{eqnarray}
where $0 \leq t \leq T$ and ${\bf f}_0$ is defined in the context of the problem discussed in
Section III. Then, it appears that if the derivative of $J^*$  with respect to ${\bf
x}=(x_1,x_2)$ exists, we can introduce a local Hamiltonian of the system as follows
\begin{eqnarray}
H= - a_0 f_0 + J_{x_1}^* x_2 + J_{x_2}^* f_2.
\label{eq34}
\end{eqnarray}
In this representation we accounted for state constraints (\ref{eq2}) and (\ref{eq3}) which in the context of problem discussed in Section III have the form (\ref{eq27}). Accounting for control constraints (\ref{eq11}) is straightforward \cite{Howlett1996A}
\begin{eqnarray}
{\cal H} = H + \lambda(u_2 - 1) + \mu( - u_2 -1), \label{eq35}
\end{eqnarray}
where $\lambda$ and $\mu$ can be identified with the Lagrangian multipliers. For this
specific case, the definition of the Hamiltonian in form (\ref{eq35}) limits the number of
degrees of freedom to two (see details of this approach in \cite{Pudney1994})
 where the objective function was taken in the form (\ref{eq23})).
However, practical applications of this approach in the context of complex dynamic systems
are limited due to the discrete nature of control  in such problems which leads to
non-existence  of derivative $\displaystyle \frac{\partial {\cal H}}{\partial u}$ in the
classical sense.  If, however, a formal operation of differentiation is performed, it is easy
to conclude that
\begin{eqnarray}
\frac{\partial {\cal H} }{\partial u_2} = \frac{\partial H}{\partial u_2} + \lambda - \mu,
\label{eq36}
\end{eqnarray}
where all the derivatives above and hereafter in the text should be understood in a
generalized sense. Under some simplifying assumptions this formal approach can be applied to
the speed control problem discussed in Section III for which the formal differentiation leads
to the following result \cite{Howlett1995,Howlett1996A}
\begin{eqnarray}
\frac{\partial {\cal H} }{\partial u} = \left \{
\begin{array}{ll}
x_2+ J_{x_2}^* + \lambda - \mu, & 0 < u_2 \leq 1,
\\[10pt]
J_{x_2}^* + \lambda - \mu, & -1 \leq u_2 <0.
\end{array}
\right.
\label{eq37}
\end{eqnarray}
In this case,  similar to our discussion in Section 4.1, further analysis can be reduced to
the consideration of five different cases, depending on the mutual location of $x_2$, $0$ and
$x_2+J_{x_2}^*$ \cite{Howlett1996A}. From a computational point of view this approach could
be efficient in computing critical speeds for automated driving strategies, but in the
general case it has the same limitations as the approach described in Section A. Of course,
in the case of simple control constraints (such as (\ref{eq11})), having the optimal velocity
$v_0(t)$, it is a standard procedure to determine the optimal control (acceleration) $u_0(t)$
by minimising the (local) Hamiltonian function. Since the velocity is strongly coupled to
control settings over the whole time interval (neither velocity nor control can be given {\em
a priori} $\forall$ $[0, T]$), practical implementation of this procedure  is problematic in
the general case. Strictly speaking, in order to determine the optimal velocity and control
globally, one needs to know the Hamiltonian which in its turns depends on those functions
\cite{Melnik1997}. However formally the Hamiltonian (or Lagrangian due to the duality
principle) can be defined locally provided the
 coupled system of equations (\ref{eq25}) and (\ref{eq27}) is solved.
Alternatively, we have to solve
 the coupled system of equations in the Hamiltonian canonical form
\begin{eqnarray}
& &
\frac{d {\bf x}^*(t)}{d t} = \frac{\partial H}{\partial { \vec\psi}} ( {\bf x}^*(t), {\bf u}^*(t), {\vec\psi}^*(t), t),
\label{eq38}
\\[10pt]
& &
\frac{d {\vec\psi}^*(t)}{d t} = - \frac{\partial H}{\partial {\bf x}} ( {\bf x}^*(t), {\bf u}^*(t), {\vec\psi}^*(t), t),
\label{eq39}
\end{eqnarray}
with the function $H$ defined as
\begin{eqnarray}
& &
H({\bf x}(t), {\bf u}(t), {\vec\psi}(t), t) \equiv f_0({\bf x}(t),
{\bf u}(t), t) +
\nonumber
\\[10pt]
& &
[ {\vec\psi}(t)]^T {\bf f}({\bf x}(t), {\bf u}(t), t).
\label{eq40}
\end{eqnarray}
Then, as follows from Theorem 4.1 for the optimality of
 control ${\bf u}^*(t)$ and  trajectory ${\bf x}^*(t)$ the
 following inequality
\begin{eqnarray}
H({\bf x}^*(t), {\bf u}^*(t), {\vec\psi}^*(t), t) \leq H({\bf x}(t), {\bf u}(t), {\vec\psi}(t), t)
\label{eq41}
\end{eqnarray}
should hold for all ${\bf u}(\cdot) \in {\cal U}$, where ${\cal U}$ is defined by
(\ref{eq4}). It is well-known that under sufficient smoothness assumptions
\cite{Kirk1970,Lions1982},  the adjoint function and the optimal performance measure are
connected by
\begin{eqnarray}
\vec\psi(t) = \frac{\partial J^*}{\partial {\bf x}} (t, {\bf x}^*(t)),
\label{eq42}
\end{eqnarray}
and hence the function $H$ in (\ref{eq40}) can be re-written in the form
\begin{eqnarray}
& &
H({\bf x}(t), {\bf u}(t), \nabla J^*, t) \equiv  f_0({\bf x}(t), {\bf u}(t), t) +
\nonumber
\\[10pt]
& &
[\nabla J^*({\bf x}(t), t)]^T {\bf f}({\bf x}(t), {\bf u}(t), t).
\label{eq43}
\end{eqnarray}
In this case the dynamics of system perturbations  (including those caused by human factors)
is accommodated formally in the derivative of the performance measure. Since  complex dynamic
systems such as ITS exhibit an intrinsic interplay between state and control aspects of the
dynamics, this accommodation does not preclude us from non-uniqueness of the solution of the
resulting model. Indeed, following \cite{Pudney1994} let $V_j$ be the speed of the vehicle at
location $x^j$, $W_j$ be the limiting speed for control setting $u^j$, and $U_j$ be the speed
at location $X_j$, where it is assumed that the speed limits are changed at distances
\begin{eqnarray}
0=X_0 < X_1 <...<X_p=X.
\end{eqnarray}
Then, triple
$(U_j, V_j, W_j)$  can be obtained by using the Lagrangian multipliers under
simplified assumptions of only three control settings, $u=1$, $u=0$,
 and $u=-1$. This triple defines  critical speeds for the interval $(X_{j-1}, X_j)$ such that $0 \leq U_j \leq V_j \leq W_j \leq M_j$ \cite{Pudney1994}.
However,   the quality of the ``speed-holding'' phase approximation by using, for example, coast-power control pairs  on each such interval depends  strongly on the number of control pairs (denoted here by $s_j$)  for this interval.
In fact, in the general case only in the limit $s_j  \rightarrow \infty$ we can obtain a unique holding speed for this interval
and to avoid undesirable vehicle speed oscillations between control switchings (e.g. between values  $V_j < M_j$ and $V_j= \min \{W_j, M_j \}$ subject to $s_j$, see the results of computational experiments in \cite{Pudney1994}).

Despite these difficulties, the problem of speed control in its general framework can be
formalised by writing down the full system of Kuhn-Tacker necessary conditions and by
including all constraints in the the globally defined generalised Lagrangian function (or
Hamiltonian, as follows from the duality principle).  Let us consider this approach in some
details. Provided ${\cal H}$ possesses sufficient smoothness, the minimisation of
 (\ref{eq35}) is a standard problem in optimization theory, and the necessary
 conditions of control optimality will follow
from $\displaystyle \frac{\partial H}{\partial u} =0$ \cite{Kirk1970,Sucharev1989}. This idea
is easy to apply in those cases where control constraints are given {\em a priori}  in a
relatively simple form \cite{Howlett1995}. However, addressing  the speed control problem in
the general case and accounting for  a complex dynamic interplay between state and control
constraints  is a much more difficult task. For example, in the case discussed in Section III
this problem is reducible to the following constrained optimisation problem
\begin{eqnarray}
H({\bf x}(t), {\bf u}(t), \nabla J^*, t) \rightarrow \min
\label{eq45}
\end{eqnarray}
\begin{eqnarray}
g_i(t) \leq 0, \; i=1,2,3, \quad g_i(t)=0, \; i=4,5,
\label{eq46}
\end{eqnarray}
(see (\ref{eq20}), (\ref{eq33}), (\ref{eq43})), subject to the following constraints
\begin{eqnarray}
& &
g_1(t)=u_2(t), \; g_2(t) = - u_2(t)-1, \; g_3(t) = -x_2(t),
\label{eq47}
\\[10pt]
& &
g_4(t)= x_1(T)-L, \quad g_5(t) = x_2(T).
\label{eq48}
\end{eqnarray}
Then, by  using classical Lagrangian multipliers for the equality constraints together with relaxing variables $\gamma_i^2$, $i=1,2,3$ for the inequality constraints, we can define the generalised Lagrangian function in the following form
\cite{Sucharev1989}
\begin{eqnarray}
L({\bf x}(t), {\bf u}(t), t, \vec\lambda, \vec\gamma) =H + \sum_{i=1}^3 \lambda_i [g_i(t) + \gamma_i^2 ] + \sum_{i=4}^5 \lambda_i g_i(t),
\label{eq49}
\end{eqnarray}
where vector $\vec\lambda$ is the vector of Lagrangian multipliers. The Kuhn-Tacker necessary conditions for the extremum of this function are
\begin{eqnarray}
\frac{\partial L}{\partial t} =0, \quad \frac{\partial L}{\partial u_2}=0, \quad \frac{\partial L}{\partial x_i}=0, \; i=1,2,
\label{eq50}
\end{eqnarray}
\begin{eqnarray}
\frac{\partial L}{\partial \lambda_i} =0   \Longleftrightarrow
g_i(t) \leq 0, i=1,2,3, \; \&  \; g_i(t) =0, \; i=4,5,
\label{eq51}
\end{eqnarray}
\begin{eqnarray}
\frac{\partial L}{\partial \gamma_i}=0 \Longleftrightarrow \lambda_i g_i(t)=0, \; i=1,2,3,
\label{eq52}
\end{eqnarray}
\begin{eqnarray}
\mbox{and, finally} \quad \lambda_i \geq 0, \; i=1,...,5.
\label{eq53}
\end{eqnarray}
In order to find the  solution of the speed control problem we should
solve  coupled system
(\ref{eq50})--(\ref{eq53}) with respect to unknown variables, $x_1, x_2, u_2, \lambda_i$.  Since
control cannot be found globally based on a locally defined velocity
function, this system should be solved in an adaptive manner. Note
 that system (\ref{eq50})--(\ref{eq53}) can be simplified substantially in some special cases, for example when
local (rather than global) solutions are sought and/or $s$ in the state equation (\ref{eq8})
is a linear function of control \cite{Howlett1990,Howlett1995}. Such simplified
considerations allow us to reduce the analysis of the Hamiltonian/Lagrangian to a small
number of control setting, as it has been explained earlier in this section. Attempts to
apply such methodologies to the general speed control problem are confronted with serious
difficulties. These difficulties are convenient to explain  at the computational level  for
the problem from section III.

Consider a trajectory of the vehicle with $n$ distinct phases
\begin{eqnarray}
P_i =(x^i, x^{i+1}), \quad i=0,1,...,n-1, \quad x^0=0, \quad x^n=X,
\label{eq54}
\end{eqnarray}
each with certain speed limits $M_{j+1}$, and the number $s_{j+1}$ of control pairs
 inside of each speed limit interval $(X_{j}, X_{j+1})$ (for example, ``coast-power''
 pairs to approximate the speed-holding phase, etc as argued, e.g., in \cite{Pudney1994})
\begin{eqnarray}
(M_{j+1}, s_{j+1}), \quad x \in (X_j, X_{j+1}), \quad j=0,1,...,p-1,
\label{eq55}
\end{eqnarray}
where $X_0=x^0$ and $X_n=x^n$.

The, we ask the following question: What values of $n$ and $p$ should be chosen to
approximate effectively the optimal trip? A simple way would be to  choose these values to
satisfy the distance and time constraints following the technique described  in
\cite{Howlett1990} (e.g. p. 468), and then to determine Lagrangian multipliers by using
methodology of \cite{Pudney1994}. However, this way cannot guarantee
 global optimality, because
  additional constraints that appear in the amended formulation of the problem
  (such as speed limits and an {\em a priori} pre-defined number of
control pairs)  should be included in the definition of the Hamiltonian (Lagrangian),  but
they are not. If we include these constraints into the definition of the
Hamiltonian/Lagrangian, the analysis cannot be reduced to only those five case
 discussed previously in this section.
In the general case, the number of speed holding phases for the entire trip can be determined
by {\em a posteriori}  estimations based on a sequential algorithm of information processing
accounting for human factors \cite{Melnik1998}. Recall that in conventional methodologies
this number is postulated {\em a priori}. At the same time, Lagrangian multipliers (see
(\ref{eq35})--(\ref{eq37})) can determine  only critical speeds within each interval
(\ref{eq55}). Algorithms for the solution of the general speed control problem can be
constructed if we take into account that the speed $V_k$, $k=1,...,n-1$ at location $x^k$ for
arbitrary (large) $n$ depends primarily on the behaviour of the system on $\Delta x_1$,...,
$\Delta x_k$, where $\Delta x_i = x^i - x^{i-1}$. We formalize this idea of the
Markovian-type controlled dynamics below by using Hamiltonian estimations. This allows us to
couple control with human-centered automation within a general mathematical framework. Note
that the model of the system as well as the objective function for the minimization as well
as constraints are subject to uncertainties as a result of dynamically changing conditions in
which the system operates.  Due to such uncertainties, the resulting control strategies may
not be optimal in the entire time interval in a classical sense. However, they are optimal
within each time interval where the same Hamiltonian estimation is used.

\subsection{Hamiltonian estimations and human-centered automation}

Effective human-centered automation is a necessary element of a
well-designed controlled intelligent transportation systems. Of course, it is not necessarily a sufficient element for an optimal performance of the overall system \cite{Kirlik1993}. However, if human factors are  incorporated into a mathematical model, then  efficient control of the overall system based on human-centered automation becomes a key tool in improving system performance.

Recall that by (\ref{eq33}) we introduced the minimum cost function. This function is based on a performance measure that allows us to include our speed control problem in a larger class of problems by considering the following functional
\begin{eqnarray}
J({\bf x}(t), t), {\bf u}(\tau): \; {t \leq \tau \leq T, \; {\bf u} \in {\cal U} } ) =
\nonumber
\\[10pt]
\int_t^T f_0({\bf x}(\tau), \tau, {\bf u}(\tau)  ) d \tau,
\label{eq56}
\end{eqnarray}
where $t$ can be any value from the closed interval $[0, T]$ and ${\bf x}(t)$ can be any
admissible state value. Now we can follow a general path of the dynamic programming approach
\cite{Kirk1970,Sucharev1989}. Since our aim is to determine the control that minimizes
(\ref{eq56}) for any admissible ${\bf x}(t)$ and for any $t \in [0,T]$, we note that the
minimum cost function for this problem can be re-written by subdividing the intervals
\begin{eqnarray}
J^*({\bf x}(t), t) = \min_{ {\bf u}(\tau) \in {\cal U}, \; t \leq \tau \leq T }
\left \{ \int_t^{t+ \Delta t} f_0 d \tau + \int_{t+ \Delta t}^T f_0 d \tau \right \}.
\label{eq57}
\end{eqnarray}
From the embedding principle used in the dynamic programming approach  (e.g. \cite{Kirk1970,Sucharev1989})
and
relationships (\ref{eq33}), (\ref{eq57}) we have
\begin{eqnarray}
J^*({\bf x}(t), t) & = & \min_{ {\bf u} (\tau) \in {\cal U}, \; t \leq \tau \leq t + \Delta t }
\left \{ \int_t^{t+ \Delta t} f_0 d \tau +
\right.
\nonumber
\\[10pt]
& &
\left.
J^*({\bf x}(t+ \Delta t), t+ \Delta t) \right \},
\label{eq58}
\end{eqnarray}
where  $ J^*({\bf x}(t+ \Delta t), t+ \Delta t)$ is the minimum cost of
the trip for the time interval $t+ \Delta t \leq \tau \leq T$ with
the ``initial'' state ${\bf x}(t+ \Delta t)$.
Applying now formally Taylor's series expansion in (\ref{eq58}) about point $({\bf x}(t), t)$, we have:
\begin{eqnarray}
& &
J^*({\bf x}(t), t) = \min_{ {\bf u} (\tau) \in {\cal U}, \; t \leq \tau \leq t+ \Delta t }
\left \{ \int_t^{t+ \Delta t} f_0 d \tau +
\right.
\nonumber
\\[10pt]
& &
\left.
J^*({\bf x}(t), t) +
\left ( \frac{\partial J^*}{\partial t} ({\bf x}(t), t) \right ) \Delta t
+ \left [  \frac{\partial J^*}{\partial {\bf x}} ({\bf x}(t), t) \right ]^T \times
\right.
\nonumber
\\[10pt]
& &
\left.
[{\bf x}(t+\Delta t) - {\bf x}(t) ] + o(\Delta t) \right \}.
\label{eq59}
\end{eqnarray}
Then, using the main property of the Landau symbol, taking into account equation (\ref{eq2})
 for $\Delta t \rightarrow 0^+$, and dividing (\ref{eq59}) by ${\Delta} t$,
 we obtain
\begin{eqnarray}
0 =  \frac{\partial J^*}{\partial t} ({\bf x}(t), t) +
\min_{{\bf u}(t) \in {\cal U} } \biggl \{ f_0({\bf x}(t), {\bf u}(t), t) +
\biggr.
\nonumber
\\[10pt]
\biggl.
\left [  \frac{\partial J^*}{\partial {\bf x} } ({\bf x}(t), t) \right ]^T
[ {\bf f} ({\bf x}(t), {\bf u}(t), t) ]  \biggr \}.
\label{eq60}
\end{eqnarray}
Further, it is easy to see that if we set $t=T$ in (\ref{eq33}) we get
\begin{eqnarray}
J^*({\bf x}(T), T) =0.
\label{eq61}
\end{eqnarray}
If we now define the system Hamiltonian by (\ref{eq43}) and take into account that  the optimal minimising control depends on ${\bf x}$,
$\displaystyle \frac{\partial J^*}{\partial {\bf x}}$ and $t$
\begin{eqnarray}
{\cal H} \left [ {\bf x}(t), {\bf u}^* \left ( {\bf x}(t), \frac{\partial J^*}{\partial {\bf x}}, t \right), \frac{\partial J^*}{\partial {\bf x}}, t \right ] =
\nonumber
\\[10pt]
\min_{ {\bf u}(t) \in {\cal U} } {\cal H} \left ( {\bf x}(t), {\bf u}(t), \frac{\partial J^*}{\partial {\bf x}}, t \right ),
\label{eq62}
\end{eqnarray}
we arrive at a mathematical model for the speed control problem (\ref{eq1})--(\ref{eq4})
represented in the form of the Hamilton-Jacobi-Bellman equation
\begin{eqnarray}
0 = \frac{\partial J^*}{\partial t} ({\bf x}(t), t) + {\cal H}  \left [ {\bf x}(t), {\bf u}^* \left ( {\bf x}(t), \frac{\partial J^*}{\partial {\bf x}}, t \right), \frac{\partial J^*}{\partial {\bf x}}, t \right ]
\label{eq63}
\end{eqnarray}
 and supplemented by condition (\ref{eq61}). The solution to this problem is understood in the
generalized sense \cite{Lions1982,Melnik1997}. Naturally, it cannot be reduced to the five
cases discussed before. Instead,  a sequential (in real time) algorithm is required to
incorporate human factors into the model via sequential  estimations of a time-perturbed
Hamiltonian approximation \cite{Melnik1997,Melnik1998}
\begin{eqnarray}
\tilde{\cal H} = \min_{u(t) \in {\cal U}, \; \; t \leq \tau \leq t + \Delta t} {\cal H}
\end{eqnarray}
for each time subinterval  $t \leq \tau \leq t + \Delta t$ with $t \in [0, T]$. This
formulation is more general than those resulted from conventional methodologies. Indeed, at
each time subinterval the Hamiltonian is allowed to change based on the information
accumulated up to that point to reflect dynamic changes in the environment in which the
system operates. A new Hamiltonian estimation should be provided based on that information.
In the context of intelligent transportation systems, Hamiltonian estimations can be obtained
efficiently with the assistance of on-board equipment, including sensors, receivers,
actuators, inter-vehicle communication systems, and on-board computers. Computational
methodologies for solving the problem for each Hamiltonian estimation are known as they were
developed for the numerical solution of HJB-based models (e.g.,
\cite{Ishii1987,Osher1988,Falcone1994,Milner1996}). Finally, a general approach to the
analysis of such models (obtained via a decision making process with limited information
available) using tools of information theory and the Markov chain approximation method can be
found in  \cite{Melnik1997,Melnik1998}.


\section{Concluding Remarks}

Efficiency of models describing the dynamics of complex systems in general, and intelligent
transportation systems in particular, often depends upon allowance made for human (e.g.,
operators/drivers) capabilities and/or limitations of these systems. As a result, the
integration of human-related design and support activities in the engineering of complex
systems become important topics of research as exemplified here on the ITS theory and
practice. In this contribution we demonstrated how human factors can be effectively
incorporated
 into the analysis and control of complex systems. As an example,
 the problem of ITS speed control, considered here as a decision making process with limited
information available, was cast mathematically in the general framework of control problems
and treated in the context of dynamically changing environments where control is coupled to
human-centered automation.  We demonstrated that the problem can be reduced to a set of
Hamilton-Jacobi-Bellman equations where human factors are incorporated via estimations of the
system Hamiltonian.
 These estimations can be obtained with the use
of on-board equipment like sensors/receivers/actuators, in-vehicle communication devices,
etc. The proposed methodology provides a way to integrate human factor
 into the solving process of the models for other complex dynamic systems.







\section*{Acknowledgements}
This work was originally inspired by discussions with members of the Scheduling
and Control Group at the University
 of South Australia, and the idea was developed further in discussions with
  the colleagues at the Mads Clausen Institute at
the University of Southern Denmark.  The author thank all of them for stimulating
discussions and a creative multi-disciplinary environment.

\nocite{*}




\begin{thebibliography}{99}

\setlength{\itemsep}{-1mm}



\bibitem{Barthelemy2002} Barthelemy,  J.P., Bisdorff, R., and Coppin, G.,
Human centered processes and decision support systems,  {\em European Journal of
Operational Research}, {\bf 136(2)}, 233--252,  2002.


\bibitem{Butts1999} Butts, K.R., Sivashankar, N. and Sun, J., Application of $L_1$ optimal control to the engine idle speed control problem, {\em IEEE Trans. Control Systems Technology}, {\bf 7}, 1999, 258--270.

\bibitem{Carroll2003} Carroll, J.M. (Ed.), HCI Models, Theories, and Frameworks. Toward a Multidisciplinary Science, Morgan Kaufmann Publishers, 2003.

\bibitem{Endo1999A} Endo, S. et al, Simulation of speed control in acceleration mode of a heavy-duty vehicle, {\em JSAE Journal}, {\bf 20}, 1999, 81--86.

\bibitem{Endo1999B} Endo, S. et al, A study on speed control law for automated driving of heavy-duty vehicles considering acceleration characteristics, {\em JSAE Journal}, {\bf 20}, 1999, 331--336.

\bibitem{Endo2000}  Endo, S. et al, A study on speed control law for automated driving of heavy-duty vehicles, {\em JSAE Journal}, {\bf 21}, 2000, 47--52.


\bibitem{Falcone1994} Falcone, M. and Ferreti, R., Discrete time high high-order schemes for viscosity solutions of HJB equations, {\em Numer. Math}, {\bf 67}, 315--344, 1994.


\bibitem{Fontaras2007} Fontaras, G. and Samaras, Z., A quantitative analysis of the European Automakers' voluntary commitment to reduce CO2 emissions from new passenger cars based on independent experimental data  • ARTICLE
{\em Energy Policy}, {\bf 35(4)}, 2239--2248, 2007.

\bibitem{Gollu1998} Gollu, A. and Varaiya, P., SmartAHS: a simulation framework for automated vehicles and highway systems, {\em Math. Comput. Modelling}, {\bf 27}, 103--128, 1998.

\bibitem{Goodrich2000} Goodrich, M.A. and Boer, E.R., Designing human-centered automation: trade-offs in collision avoidance system design, {\em IEEE Trans. ITS}, {\bf 1}, 2000.


\bibitem{Herty2006} Herty, M. and Rascle, M., Coupling conditions for a class of
second-order models for traffic flow, {\em SIAM J. on Mathematical Analysis}, {\bf 38 (2)},
595-616, 2006.

\bibitem{Howlett1995}
Howlett, P.G. and Pudney, P.J.
{\em Energy-Efficient Train Control,}
 Springer-Verlag, 1995.

\bibitem{Howlett1990} Howlett, P., An optimal stategy for the control of a train, {\em J. Austral. Math. Soc.: B}, {\bf 31}, 1990, 457--471.

\bibitem{Howlett1996} Howlett, P., Optimal strategies for the control of a train, {\em Automatica}, {\bf 32}, 519--532, 1996.


\bibitem{Howlett1996A} Howlett, P.G., Personal communication, 1996.

\bibitem{Ishii1987} Ishii, H., Perron's method for Hamilton-Jacobi equations, {\em Duke Mathematical Journal}, {\bf 55}, 369--384, 1987.

\bibitem{Kesseler2006} Kesseler, E. and  Knapen, E.G., Towards human-centrednext term design: Two case studies, {\em Journal of Systems and Software}, {\bf 79(3)}, 301--313, 2006.

\bibitem{Kiencke2006}  Kiencke, U., Nielsen, L., Sutton, R., et al.,
The impact of automatic control on recent developments in transportation and vehicle systems,
 {\em Annual Reviews in Control}, {\bf 30 (1)}, 81-89, 2006.

\bibitem{Kirk1970} Kirk, D.E., Optimal Control Theory, Eglewood, N.J., Prentice Hall,  1970.

\bibitem{Kirlik1993} Kirlik, A., Miller, R. A. and Jagacinski, R. J., Supervisory control in a dynamic and uncertain environment: a process model of skilled human-environment interaction, {\em IEEE Trans. on Systems, Man, and Cybernetics}, {\bf 23}, 1993, 929--951.

\bibitem{Kolmanovsky2000} Kolmanovsky, I., van Nieuwstadt, M. and Sun, J., Optimization of complex powertrain systems for fuel economy and emissions, {\em Nonlinear Analysis: RWA}, {\bf 1}, 2000, 205--221.


\bibitem{Kraiss2001} Kraiss, K.-F. and Hamacher, N., Concepts of user centered automation,
 {\em Aerospace Science and Technology}, {\bf 5 (8)}, 505--510, 2001.

\bibitem{Lions1982} Lions, P.-L., Generalized Solutions of Hamilton-Jacobi Equations, Pitman, Boston, 1982.


\bibitem{Manzie2007} Manzie, C., Watson, H., and Halgamuge, S.,   Fuel economy improvements for urban driving: Hybrid vs. intelligent vehicles  • ARTICLE
{\em Transportation Research Part C: Emerging Technologies}, to appear 2007.


\bibitem{Mayer2006} Mayer, F. and Stahrea, J., Human-centred systems engineering (Editorial to a special issue),
{\em Annual Reviews in Control}, {\bf 30(2)}, 193--195, 2006.



\bibitem{Melnik1997} Melnik, V.N., On consistent regularities of control and value functions, {\em Numer. Funct. Anal. and Optimiz.}, {\bf 18}, 401--426, 1997.

\bibitem{Melnik1998} Melnik, R.V.N., Dynamic system evolution and Markov chain approximation, {\em Discrete Dynamics in Nature and Society}, {\bf 2}, 7--39, 1998.

\bibitem{Milner1996} Milner, F.A. and Park, E.-J., Mixed finite-element methods for HJB-type equations, {\em IMA J. Numer. Anal.}, {\bf 16}, 399--412, 1996.

\bibitem{Osher1988} Osher, S. and Seithian, J.A., Fronts propagating with curvature-dependent speed: algorithms based on Hamilton-Jacobi formulations, {\em J. Comput. Physics}, {\bf 79}, 12--49, 1988.


\bibitem{Pontryagin1986}
Pontryagin, L.S. et al,
{\em The Mathematical Theory of Optimal Processes,} Gordon \& Breach, 1986.

\bibitem{Pudney1994} Pudney, P. and Howlett, P., Optimal driving strategies for a train journey with speed limits, {\em J. Austral. Math. Soc. Ser. B}, {\bf 36}, 38--49, 1994.


\bibitem{Qi2005}  Qi, Y and  Zhao, YYJ,
Energy-efficient trajectories of unmanned aerial vehicles flying through thermals, {\em J. of
Aerospace Engineering}, {\bf 18 (2)}, 84-92, 2005.

\bibitem{Rouse1993} Rouse, W. B., Edwards, S.L. and Hammer, J.M., Modeling the dynamics of mental workload and human performance in complex systems,  {\em IEEE Trans. on Systems, Man, and Cybernetics}, {\bf 23}, 1993, 1662--1671.


\bibitem{Seto1999} Seto, Y. and Inoue, H., Development of platoon driving in AHS,  {\em JSAE Journal}, {\bf 20}, 1999, 93--99.


\bibitem{Shahar2006} Shahar, Y. et al,  Distributed, intelligent, interactive
visualization and exploration of time-oriented clinical data
and their abstractions, {\em Artificial Intelligence in Medicine }, {\bf 38(2)}, 115--135, 2006.


\bibitem{Sivashankar1999} Sivashankar, N. and Sun, J., Development of model-based computer-aided engine control systems, {\em Int. J. Vehicle Design}, {\bf 21}, 1999, 325--343.


\bibitem{Stotsky1995} Stotsky, A., Chien, C.-C. and Ioannou, P., Robust platton-stable controller design for autonomous intelligent vehicles, {\em Math. Comput. Modelling}, {\bf 22}, 1995, 287--303.

\bibitem{Sucharev1989} Sucharev, A., Timochov, A., Fedorov, V., {\em Optimization Methods}, Nauka,  1989.


\bibitem{Wren2006} Wren, C.R., Minnen, D.C., and Rao, S.G.,  Similarity-based analysis for large networks of ultra-low resolution sensors,
{\em Pattern Recognition}, {\bf 39(10)}, 2006, 1918-1931.

\bibitem{Zhuan2006}  Zhuan, X, and Xia, X.,
Cruise control scheduling of heavy haul trains, {\em  IEEE Trans. on Control Systems
Technology}, {\bf 14(4)}, 757-766, 2006.

\end{thebibliography}
\end{document}